\documentclass[12pt]{article}
\usepackage{graphicx}
\begin{document} 
\baselineskip 15pt

\bigskip
\centerline{\bf Urban and Scientific Segregation: The Schelling-Ising Model}

\bigskip
Dietrich Stauffer and Christian Schulze\\
Institute for Theoretical Physics, Cologne University\\D-50923 K\"oln, Euroland

\bigskip
Abstract: Urban segregation of different communities, like blacks and whites
in the USA, has been simulated by Ising-like models since Schelling 1971. This
research was accompanied by a scientific segregation, with sociologists and
physicists ignoring each other until 2000. We review recent progress and also
present some new two-temperature multi-cultural simulations.

\section{Introduction}
The Schelling-Ising model of urban segregation is for two reasons of interest 
for readers of sociophysics \cite{fortunato} papers: 1) It may explain aspects 
of racial, ethnic, religious, ... segregation in large cities (``ghetto 
formation''); 2) it is an example of decades-long non-cooperation between social
sciences and physics.

Harlem in New York City, USA, is the ``black'' residential district of 
Manhattan, where hundreds of thousands of Afro-Americans live. In most of the
rest of Manhattan the population is at least 80 percent ``white'' (ignoring
Hispanic), while Harlem is at least 80 percent ''black'' \cite{sethi}. Similar
residential segregation has been observed in other places, along ethnic,
religious or other barriers, though often on a smaller scale. The city of the 
present authors \cite{friedrichs} has since years a tenth of its population 
from Turkish background and now it tries to deal with a plan to build a large 
mosque here, after years of partial segregation into several ``Turkish'' city 
districts.

Such segregated residential districts can be caused by external forces, like
the orders of Nazi Germany that all Jews have to live in small ghettos
of Warsaw, Cracow, .... The alternative to be discussed here is the emergence of
urban segregation without external force, only through self-organisation due
to the wishes of the residents. 

This second possibility was already pointed out 25 centuries ago by the Greek
philosopher Empedokles, who (according to J. Mimkes) found that humans are like
liquids: some live together like wine and water, and some like oil and water
refuse to mix. This idea was formulated more clearly by the German poet Johann
Wolfgang von Goethe in ``Wahlverwandtschaften'' two centuries ago (also 
according to J. Mimkes). The rest of this article deals with the implementation 
of this basic idea. 

Schelling published in 1971 the first quantitative model for the emergence 
of urban segregation \cite{schelling}, in the same year in which physicist 
Weidlich \cite{weidlich} published his first paper on sociodynamics. Schelling
in 2005 got the economics ``Nobel'' prize; his 1971 paper has an 
exponentially rising citation rate and is the second-most-cited paper in this
Journal of Mathematical Sociology. His model is a complicated zero-temperature
Ising ferromagnet but this similarity to statistical physics was overlooked
\cite{aydin}. Only very recent publications \cite{kirman,solomon} pointed out
that Schelling's original model fails to form large ghettos and leads only 
to small clusters of residences for the two groups in the population. Long 
before, Jones \cite{jones} introduced some randomness into the model to produce
large ghettos; this good paper, in turn, was mostly ignored for two decades. 

The first citation to Schelling 1971 from physicists known to us is the book
of Levy, Levy and Solomon \cite{levy}, nearly three decades after publication.
One of us (DS) learned about the Schelling model from a meeting on Simulating
Society in Poland, fall 2001 (though he should have learned it from \cite{levy}
of which he got preprints). He then was advised in spring 2002 by Weidlich
\cite{weidlich} to study learning for urban segregation, and forwarded 
this idea to \cite{ortmanns,schulze}. Vinkovic and Kirman \cite{kirman}
reviewed nicely the physics of hundred years ago in the Empedokles-Goethe sense,
but ignored the physics research of recent decades. We are not aware of 
sociology to have taken note of the Ising model in the follow-up of the
Schelling model \cite{schelling}; as mentioned sociologists also mostly
ignored the paper of their colleague Jones \cite{jones} in a sociology 
journal.

Thus we see here not only residential segregation, but also scholarly 
segregation, with physicists ignoring the Schelling model until recently,
and sociologists ignoring the similarity of the Schelling to the Ising model 
until now. Books like the present one can help to bridge the gap. 

This article is written for physicists; an earlier article \cite{solomon} was 
supposed to help also non-physicists understand the Ising model in connection
with urban segregation. 

\section{Schelling model}

In order to check whether urban segregation can emerge from personal preferences
without any overall management, discrimination etc, Schelling published in
1971 a modification of the spin 1/2 Ising model at zero temperature. The 
majority of sites on a square lattice are occupied by people who belong to
one of two groups A and B. The initial distribution is random: A and B in equal 
proportion; a minority of sites is empty. Everybody likes to be surrounded by 
nearest and next-nearest neighbours of the same kind. People are defined as 
unhappy if the majority of their neighbours belongs to the other group, and 
as happy if at least half of these neighbours belong to the own group. (Empty
sites do not count in this determination.) At each iteration with random 
sequential updating, unhappy people move to the nearest place where they can
be happy \cite{schelling}.

As a result, clusters are formed where A residences stick together, and also 
B people cluster together. These clusters remain also when details of the 
model, like the thresholds of happiness, are changed \cite{schelling}. The 
lattices studied (by hand !) in that paper were too small to indicate whether
the clusters become infinite in an infinite lattice, i.e. whether phase 
separation like oil in water happens. Only a third of a century later, a
computational astrophysicists and an economist \cite{kirman} showed that the 
clusters
remain small even if large lattices are simulated. Thus the original Schelling
model is unable to explain the formation of large ghettos like Harlem, but
it can explain the clustering of a few B residences surrounded by A houses. 
We have not yet found papers from sociology pointing out this failure, though
Jones \cite{jones} may have noticed it when he improved the model (see below),
without stating it explicitly in his publication. 

Even without any computer simulation one can guess that the original Schelling
model had troubles. Solomon \cite{solomon} pointed out that the following
B cluster will never dissolve from within: Each B has four or more B sites 
in its neighbourhood of eight, feels happy and has no intention to move. 
The A have even less reason to move.
\medskip

A A A A A A A A

A A A A A A A A

A A A B B A A A

A A B B B B A A

A A B B B B A A

A A A B B A A A

A A A A A A A A

A A A A A A A A

\medskip
Also horizontal strips of A and B sites lead to blocking, as in Ising
models \cite{redner}, since even at the interface everybody is happily 
surrounded by neighbours mostly of the own group.
To avoid such blocking of cluster growth, Jones \cite{jones} removed a small 
fraction of people randomly, and replaced them by people who feel happy in these
vacancies. Then really large ghettos are formed. (The earlier Dethlefsen-Moody 
model also involves randomness and still gives only finite clusters, though 
larger than in the Schelling version, according to M\"uller \cite{muller}.) 

An alternative way to introduce some randomness \cite{kirman} is to allow
people to move even when their status (happy or unhappy) is not changed by
this move. Then domains were simulated to grow towards infinity, and the 
detailed behaviour of this growth process was studied later \cite{asta}.
This assumption seems at first sociologically nonsensical: Why should anyone
go through the troubles to move if this does not improve the situation? 

However, one may regard such ``useless'' moves as coming from forces external
to the model, like when one moves to another city in order to switch employment.
In that case, however, one may also be forced from a happy place to one where
one feels unhappy, as is done below through a finite temperature $T$. Thus the
model of \cite{kirman,asta} is a nice physics model with limited sociological 
appeal.

[Schelling assumed that A people stay A people forever, and the same applies 
to B people. This may not be true with respect to religion or citizenship but
is correct with respect to skin colour. However, people also move out into 
another city, or in from another city, and thus within the simulated area the
number of A and B people can change. The blacks in Harlem may have had ancestors
who worked on tobacco or cotton plantations, but these plantations were further
south and not in Manhattan. Jones \cite{jones} already simulated fluctuating
compositions, and in contrast to an assertion in \cite{kirman}, the Ising 
models has been  simulated since decades for both fixed (Kawasaki) and 
fluctuating (Glauber, Metropolis) magnetisations.]

\section{Two groups, one temperature}

\subsection{Schelling and Ising at positive $T$}

Life is unfair, and we do not always get what we want. Thus for accidents 
outside the model, like loss of job, marriage, ..., we sometimes have to 
change residences even if we like the one in which we live. Thus we may 
not only have neutral moves as in \cite{kirman} from unhappy to unhappy or from 
happy to happy, but also from happy to unhappy with some low probability.
This is the basis of thermal Monte Carlo methods (Metropolis, Glauber, 
Heat Bath algorithm), also for optimization (simulated annealing etc.) 
\cite{schneider}, where the energy (unhappiness) is increased by $\Delta E$ 
with a probability $\propto \exp(-\Delta E/k_BT)$. And then it also makes 
sense to look at different degrees of (un)happiness, that means to treat
the number of neighbours from the other group as $\Delta E$. The standard
two-dimensional Ising model then is the simplest choice. 

In this sense, $T$ plays the strength of the external noise which pushes us
to move against our personal preferences. Instead, the temperature $T$ can also
be interpreted as 
``tolerance'' \cite{mimkes}: The higher $T$ is, the more are we willing to live
with neighbours of the other group. In the limit of infinite $T$ the composition
of the neighbourhood plays no role for changing residences; in the opposite
limit of zero $T$ we never move from happy to unhappy in the Schelling version;
and never from a smaller to a larger number of ``wrong'' neighbors in the 
Ising version. In the Ising version, again one can work with a constant or 
a fluctuating composition of the total population.

In the Schelling version at finite $T$, with probability exp($-1/T$) we 
consider moving out of a site where we are happy, and we always consider 
moving out from a site where we are unhappy \cite{solomon}. If we consider
moving, we check empty sites in order of their distance from us, and accept the
new site if we are happy there. Otherwise we accept it only with probability 
exp($-1/T$) if we are unhappy there, and instead continue to look for empty 
sites further away with probability $1-\exp(-1/T)$. Then 
large ghettos are formed though very slowly \cite{solomon}. 
Similar results are found in various variants \cite{muller}.

In the Ising version at finite $T$ we have much simpler and clearer definitions
and flip a spin (change the group at one site) with probability $1/[1+\exp(
\Delta E/k_BT)]$. The degree $\Delta E/J$ of unhappiness is the number
of neighbours of the other group, minus the number of neighbours of the same 
group. Thus, for example, neutral moves \cite{kirman} are made with probability
1/2, and moves from four neighbors of the own group to four neighbours of the
other group only with probability exp$(-8J/k_BT)$ at low $T$. (The 
proportionality constant $J$ is called exchange energy in physics.) 
This Ising model shows growth of infinitely large domains (``ghettos'') for
$T < T_c$ and only finite clusters for $T > T_c$ with $J/k_BT_c = \ln(1 + 
\sqrt 2)/2$, as is known since two-thirds of a century. Such pictures are
also nice for teaching \cite{isinggould}.

More novel Ising simulations \cite{ortmanns} took into account the ``learning'':
Through education etc people become more tolerant of others or more similar to 
others. Thus Meyer-Ortmanns \cite{ortmanns} showed how at low tolerance $T$ 
compact ghettos are formed, which dissolve through Kawasaki kinetics (constant 
composition of the population) if $T$ is suddenly increased. It does not 
matter whether this learning comes from the groups becoming more similar to each
other (decrease of $J$) or from both groups increasing their tolerance $T$: 
Only the ratio $J/k_BT$ enters the simulations. We will return to this learning 
when dealing
with more than two groups in a Potts-like segregation model \cite{schulze}.

V.Jentsch and W.Alt at Bonn University's complexity centre suggested to have 
an individual $T(i)$ for each different site $i$, and to introduce a feedback:
If one sees that all neighbours are of the own group, one realizes that
strong segregation has happened, does not like this effect, and thus increases
the own $T(i)$ by 0.01. If, in contrast, all neighbours belong to the other 
group, one also dislikes that and decreases the own $T(i)$ by 0.01. Also,
one forgets the tolerance one has learned this way and decreases $T(i)$ at
each time step by a fraction of a percent. Then \cite{muller} depending on this
forgetting rate, a spontaneous magnetisation remains, or it goes down to zero,
while the final self-organized average $T$ may differ only slightly in these two
cases. 

\subsection{Money and random-field Ising model}

\begin{figure}[hbt!]
\begin{center}
\includegraphics [angle=-90,scale=0.35]{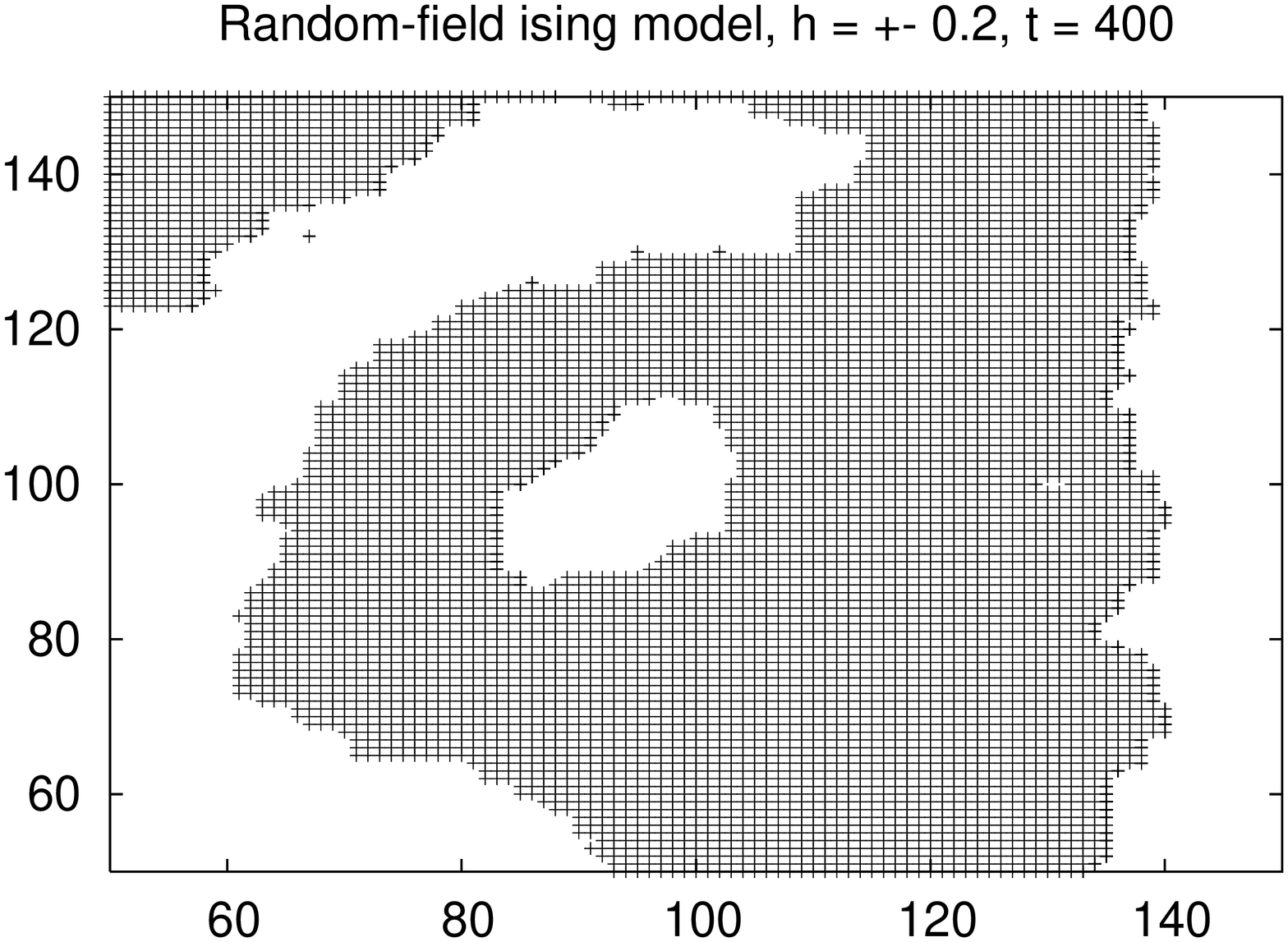}
\includegraphics [angle=-90,scale=0.35]{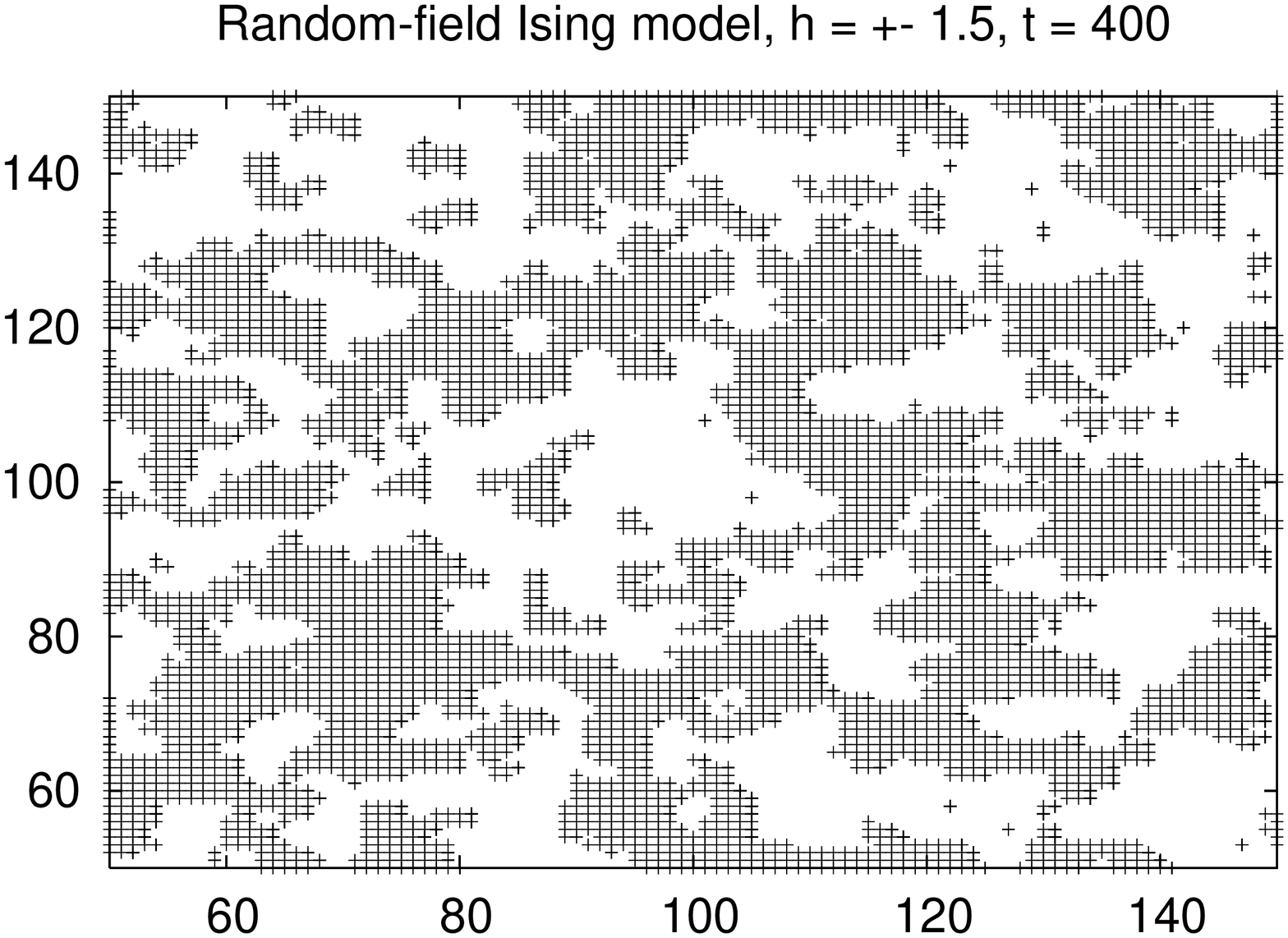}
\end{center}
\caption{Large domains (upper part) and small clusters (lower part) for small
(0.2) and large (1.5) random fields, respectively, at $T = 1$. Field and
temperature are measured in units of $J/k_B$. After \cite{sumour}.
}
\end{figure}

Often, a population can be approximately divided into rich and poor. Starving
associate professors and luxuriously living full professors are one example,
but poorer immigrants and richer natives are more widespread. Poor people 
cannot afford to live in expensive houses, but rich people can. If there are
whole neighbourhoods of expensive and cheap housing, then these housing 
conditions enforce a segregation of rich from poor, and this segregation does
not emerge in a self-organised way. The more interesting case allows for 
self-organisation of domains by assuming that each residence randomly is either
expensive or cheap, with no spatial correlations in the prices. Does this
model lead to spatial correlations between the two groups, assumed as poor
and rich?

A suitable model is the random-field Ising magnet, where each site of a
square lattice carries a magnetic field $\pm h$ which is randomly up (expensive)
or down (cheap). The resulting addition to the energy prefers up spins on 
the expensive and down spins on the cheap sites. Simulations of the asymptotic
behaviour are difficult \cite{hartzstein}, but are not needed for urban 
segregation happening in finite times on finite samples \cite{sumour}. Fig.1
shows two examples: A small random field allows for large domains, upper part,
while a large random field allows only small clusters, lower part. Sumour 
et al \cite{sumour} presented at lower temperatures also a time series of 
pictures, separating growing from non-growing domains. They also listed a 
complete 50-line Fortran program.

Thus we see that the personal preferences are balanced by the random field, i.e.
by the prices of housing; spatial correlations over short distances still exist
for moderately high fields.

\section{Several groups, two temperatures}

While the separation into black and white corresponds to USA traditions, reality
often means the coexistence of more than two major groups (like Hispanics 
in the USA). Similarly, while the spin 1/2 Ising model is the most basic
and most widespread model in statistical physics, the Potts and Blume-Capel
models have generalized it to more than two possible states of each lattice
site \cite{schulze}.

Also, while equilibrium physics has one temperature $T$, in the above 
segregation studies we used $T$ for both the tolerance \cite{mimkes} and the
external noise \cite{solomon}. In reality, these two effects should be described
by two independent parameters $T_1$ for tolerance and $T_2$ for noise. And
again the tolerance can be different for different people and can self-organize
to some average tolerance through a feedback with the local amount of 
segregation, as discussed above for two states and one local temperature.

If each site $i$ of a lattice carries a variable $q_i = 1,2, \dots, Q$, 
it may correspond to $Q$ different ethnic groups. For Germany, $Q=5$ may 
represent immigrants from the Iberian peninsula ($q=1$), from Italy ($q=2$),
from the former Yugoslavia ($q=4$), from Turkey ($q=5$) and native German tribes
($q=3$). The more dissimilar these groups are, the less they like each other.
Thus the energy $E$ or unhappiness is assumed to be
$$E = J \sum |q_i -q_k| \eqno (1)$$
where the sum goes over all nearest neighbour pairs on a $L \times
L$ square lattice, with helical boundary conditions.
We measure all temperatures in units of the Potts critical temperature
[2/ln$(1+\sqrt Q)]J/k_B$, even though this model does not have a sharp 
phase transition for $Q> 2$ \cite{schulze}; only for $Q=2$ is $T=1$ at the
Curie temperature or segregation point. People exchange residences with others
anywhere in the system (except in their immediate neighbourhood), with 
Glauber probability $1/[1+\exp(\Delta E/k_BT)]$, thus the number of members
of the $Q$ groups remains the same.

As in the above Ising case \cite{ortmanns}, we allow for learning. Here that 
means we start at a low temperature 0.5 and with increasing time $t$ we increase
$T$ during a learning time $\tau$ to a higher temperature 2.5 according to 
$$T(t) = 0.5+2t/\tau \eqno (2) \quad .$$
We measure the amount of segregation through the correlation 
$$ C(t) = \langle n \rangle Q/4 \eqno (3)$$
where $n$ is the number of nearest neighbours of the same group and averaged
over all sites. (Alternatively one can average only over the sites occupied by 
the central group.) In this way, $C = 1$ for the initial random distribution
of people, and the increase of $C$ above unity measures the amount of
segregation. 

The following picture shows a small configuration at $\tau = 0$, i.e. when 
the tolerance is immediately at its maximum final value $T = 2.5$; hundred
sweeps through the lattice were made, and only the central group 3 for $Q=5$ 
is shown.

{\small
\begin{verbatim}

                *  ** * *            **         *                    *  * *   **
       *           * **                      *                          *** *   
           *     ** ****      *        *    **        *   *          *   **  *  
 * *                  ** *          *  * **** ***                   **   *    * 
   *    *  *                  *      ***********          *        **       *   
          ***        *              ***   * ****                   ** **        
  *  *    *  * **                         **  ***      *            ****        
       *  **  ****               ***           *      **        ** *  *   *   **
           * *            *       **  *        *      **       * * *****  ****  
   * *  ****  *           **      *   * **     *    ***  *        * * *         
    ***** ***                                  *     *   *    *    ** ***       
    *  *   **   *  **                    **          * ****   **      **   *    
     *                       *     *             ** ***  * **        *  *       
 *      *    **                                ****  ***                *       
   *   *  **   * **          * *                **    **     *                **
 *  **     * *     *          ***     *        *    *         *          *    **
 *   *  *                        **  **     *     *    *   **  * **         **  
                  *     *     ** *   *   **   * *            *** **        * ** 
     *                      * **     *    *                    **            *  
                  **       *  ***       *    *              **      *           
 *  **               *        **         *  **              *                   
    * *          **           *     ****                   **     *             
      *****      *   ***       **   ****                          *            *
      *** *                   ***               *        *                      
      **   ** *****      **     **    *       * *       *  *              *     
            *  *        **             *      *          **                     
   ***              *  ****                             **          * **        
    **                 * **      *     *       *        ***           **        
    ** *          *   **     *  *   * **                 **      *              
    *   ***   ** **       *   **    *         *         *   **                 *
      ****     **               **  **                      * *  **             
        **      * *               *                  *        *  *       *      
  * *   *              * *                     *      **       *        **  *  *
   *          *   *    * ***   ** *        **  **    *                ******  * 
                        * *      **                 *               * * *** **  
                **     **     **              ** ** *                **   **    
       **   **  *     **       ***  *    *       ** *                 **      * 
      ***   * **   *    *       *  **           *        *               ** *   
     **** ****  ****               *          ***                       ***   * 
     ** ** *  * ****          *   **       *    **                       * *  * 
    ** * **   * *  ***     *          *       ** *                            * 
      * *        *  *     *       **      *****                                 
         * *         *    *         *  *    ** *        *                 *     
            * *               *     * *    **                              *    
            *                      *       *** **                               
             *  *          *** *  **     *  ***                       *         
  *            **         ****     **          *        *         **  **        
            *            *  * **  *       * ****                 *    **        
                      * *    *  * *   ******** ***                     *        
                      *****     *   **  *    *   **                  ***        
   *                *           *   **  * *      ***  **              **        
                    *  *   *       **           **  *  *     *         *        
    *               *   *    *           *      * *   **** * **         *      *
   *             *   *       *                    *  **   ** **   *     *   * * 
      *  *       **  ** **      *                **  *    ***    **  *          
      *    *        **  *                               ** ***    ****  *       
       *  **      *     ****  **** **                *     *    **   *          
      *    *              *    *    *               *                *          
  *   *       *                *    *                *           *       *      
  **  * **             **   ****   **                            * *    **     *
    *                   *** *      *     *  *                    ****   *       
                     ***  ***  *           ***      *   *  *     *  **          
   ***               ****             **  ****     ***  *        **  *         *
    **               ***  *            ******     *****        *        * *  *  
            *                          * **         * **  * *  *          ****  
                **     *    *   *      *    *       **         *     * ****  * *
 *          * *    *        *        * *****                ****      ***** *   
 *        * ***          *     *  *    * ***    * ****       *   *       ***  * 
 * **  **  ***   *      *  *  ****  *  * **** **             *      *  ** *   **
  **  *** ** ****  *** ***   **     *  *    ****  **                    **      
     *****   *** **  ** **          ****  *      ****       **           * *    
   ****     ***      *           * **       **   **               *          *  
    ** **   *     ** **  *              *         **   *         *           ** 
            *   *   ****                **    **  ***                ***      * 
            ** *    ***  *              *** *  ** * *                * *      * 
  *      *  *  ****                         * *  ** * ***                       
 ***        **  * *    **            *           *      ******                  
   *    *     *       * * *       **            *      ***                      
  ***      * ** *  ***  *          * *    *            *                    *  *
\end{verbatim}
}

Now we bring the corresponding configuration for much slower learning $\tau=20$:
{\small
\begin{verbatim}

 *  ****       **  ***       ** *** ***   *****     ***             ** * *      
  ******       *** *            *******    **       *     *                     
   ******      *****     *  **   ******     *       *                           
    ****     ********         ********              *                           
           *   **               *   *                                           
                                                                                
                                                                                
                                                  *                             
         *                                   **  *                    **   ***  
        **                                   *** *                 **        *  
       ****                                *******                 **           
       ******                             ****              *   *  **           
        ******                            **                *           *       
  *      **                              **                 *                   
  *       *******                 *****                     *         **        
            *  ***              *   **                                 *        
  *                                 ** * **                 *         **        
  *                                 *   *                   *        *          
  *                                     ***                 *        *          
  *   *        *                                          ***                   
   ****  *** *** ***                                     **        *            
   **    *** *** ******        **               ** *      *                     
          ***** *******       **                   *       *     **             
          *************   ***                      *     ****                   
        **   ********** ***                        *     *****             *    
        *     ***    ** **                               ** **        *         
               *        ***                              *****                  
                         **          *                    ***                *  
                          *         **         *          ***                ** 
                          **       *                     * ***                * 
        **                **                          ********                * 
        **                *                         ***********               * 
       *         *        *                       ***********         ****   ***
             ******      **                         *******             **  ****
 *****         ****      **     *                    ******    *          ******
 ****              **   ***     *                      ***       *        ******
 **                *  *****     *                      ***        **           *
                      *****     *                      ***         **           
           **          ****     *                      ***           **         
           *****     *******   ***                     *** **  ***  ****        
           ******     ****** *                         ****    *********        
              ***     ********                         ***      *********       
             ****     ********                         ***      *********       
       *       **      *******                        ****        * *******     
        *      **     ** *****                    *    **              ****     
               **    ******      ***                                   ****     
         *     **    ***          **             *       *              ** *    
         ** *        ***           ***                    *            *****    
                       *                         **                    *****    
                 *                                        **           *****    
                                          * **    **                  **** **   
                                         *****     **                 ****  ****
    ** *                                 *******   ***                 ***     *
   *****                                   ***             *          ****      
 *******                                   ***            **           ***      
 ***                  *    ***              **            *           **********
 **                   ***  ***              ***                      ***********
 ***   *             **  * ***                            *        *************
       *             ***     *    **                      *       *   ******  **
      **    *        **           **               *      **         *****      
    ****    **                    **                *      *     *       *      
    ***    * **        *      * ** *                 *       ****               
     ***                      ********  *                    ****               
     *                   **   ***************                *****              
                          *   ****************               ****          *** *
               *                ******   * ***        *    *******          ****
               **                ****       **        *  * * *****          *   
      **       **                  *         **       * **** * ***              
               ***                 ***        *       *  ***   **               
       *         ** **             ****       **     *   ***                    
       **          ****              **       **   ***  **                      
        *            *******   *     **        *   **   **                      
        **           *******                 ***   *                            
        *            ******      *            **                                
        *               **       **    *     ***   *                            
    *   ***             **           * ** *******                 ******        
    ** ****             ***        ** **********                  *******       
    **********          ***        ***********                  ************    
    ************     *             ****** *********          *    **** ***    **
\end{verbatim}
}

The second picture shows a clear compactification compared with the first,
though the domains grow only slowly to infinity.
The situation may correspond more to Cologne, Germany, than to Harlem, New York 
City. More quantitatively, the upper part of Fig. 2 (plus signs) shows how
segregation increases to high intermediate levels if people do not learn fast 
enough.

\begin{figure}[hbt!]
\begin{center}
\includegraphics [angle=-90,scale=0.5]{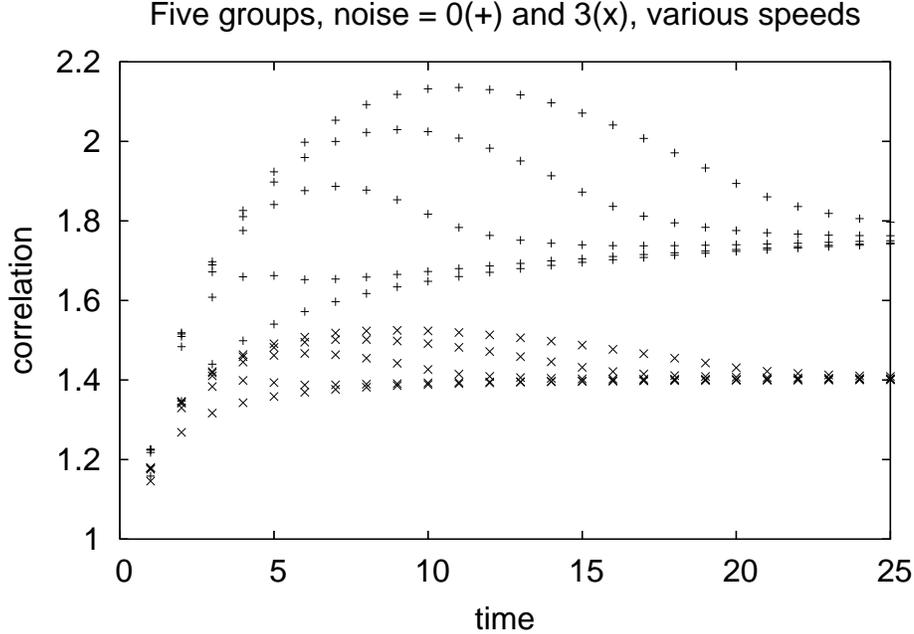}
\end{center}
\caption{Segregation for $Q=5, \; L = 30001$ at learning times 0, 5, 10, 15 
and 20, from bottom to top. The plus signs correspond to no noise, the x signs 
to strong noise $T_2=3$.
}
\end{figure}

\begin{figure}[hbt]
\begin{center}
\includegraphics [angle=-90,scale=0.4]{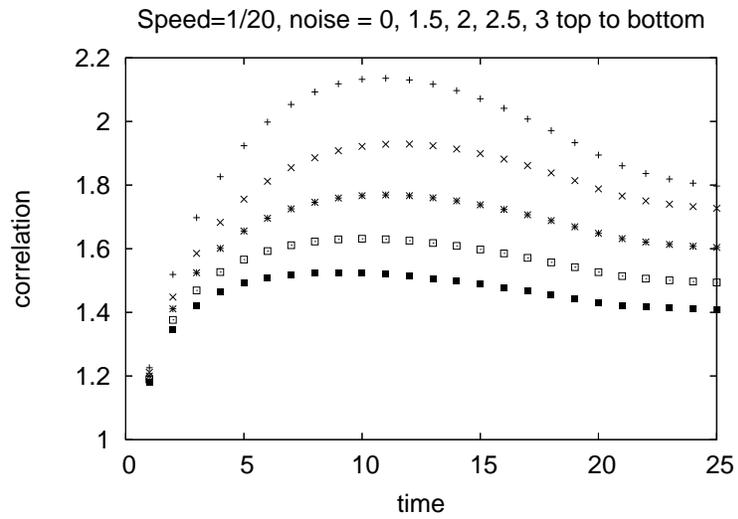}
\end{center}
\caption{As Fig.2, for learning time $\tau = 20$, and noise increasing from 
top to bottom.
}
\end{figure}

\begin{figure}[hbt!]
\begin{center}
\includegraphics [scale=0.55]{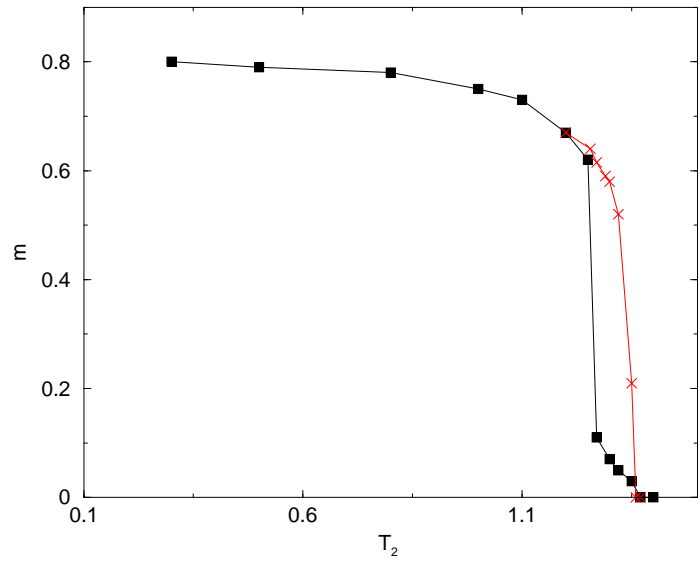}
\end{center}
\caption{First-order transition with hysteresis for \'Odor's two-temperature 
Ising model.
This and the last figure were kindly supplied by G. \'Odor for this review.}
\end{figure}

\begin{figure}[hbt!]
\begin{center}
\includegraphics [angle=-90,scale=0.3]{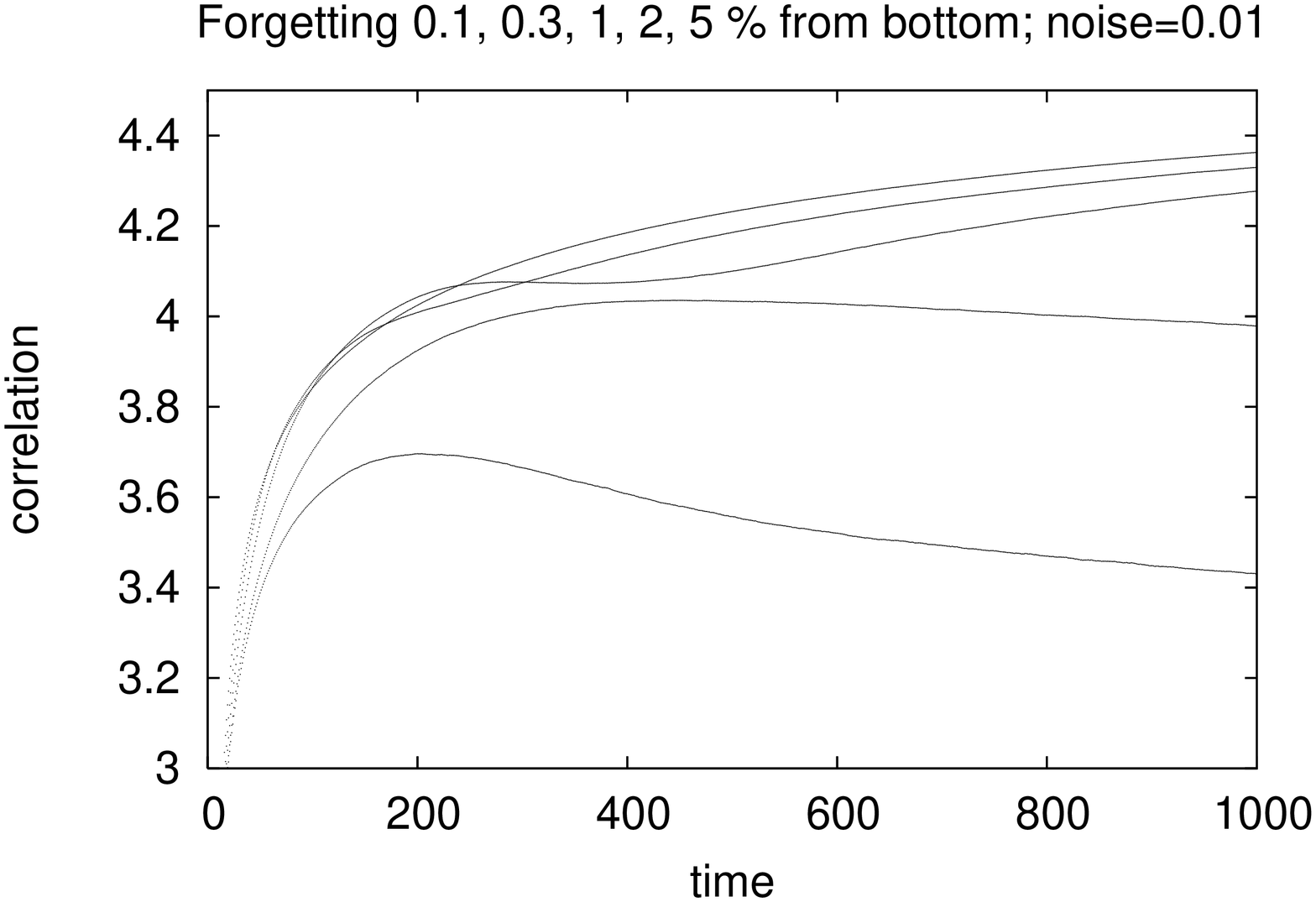}
\includegraphics [angle=-90,scale=0.3]{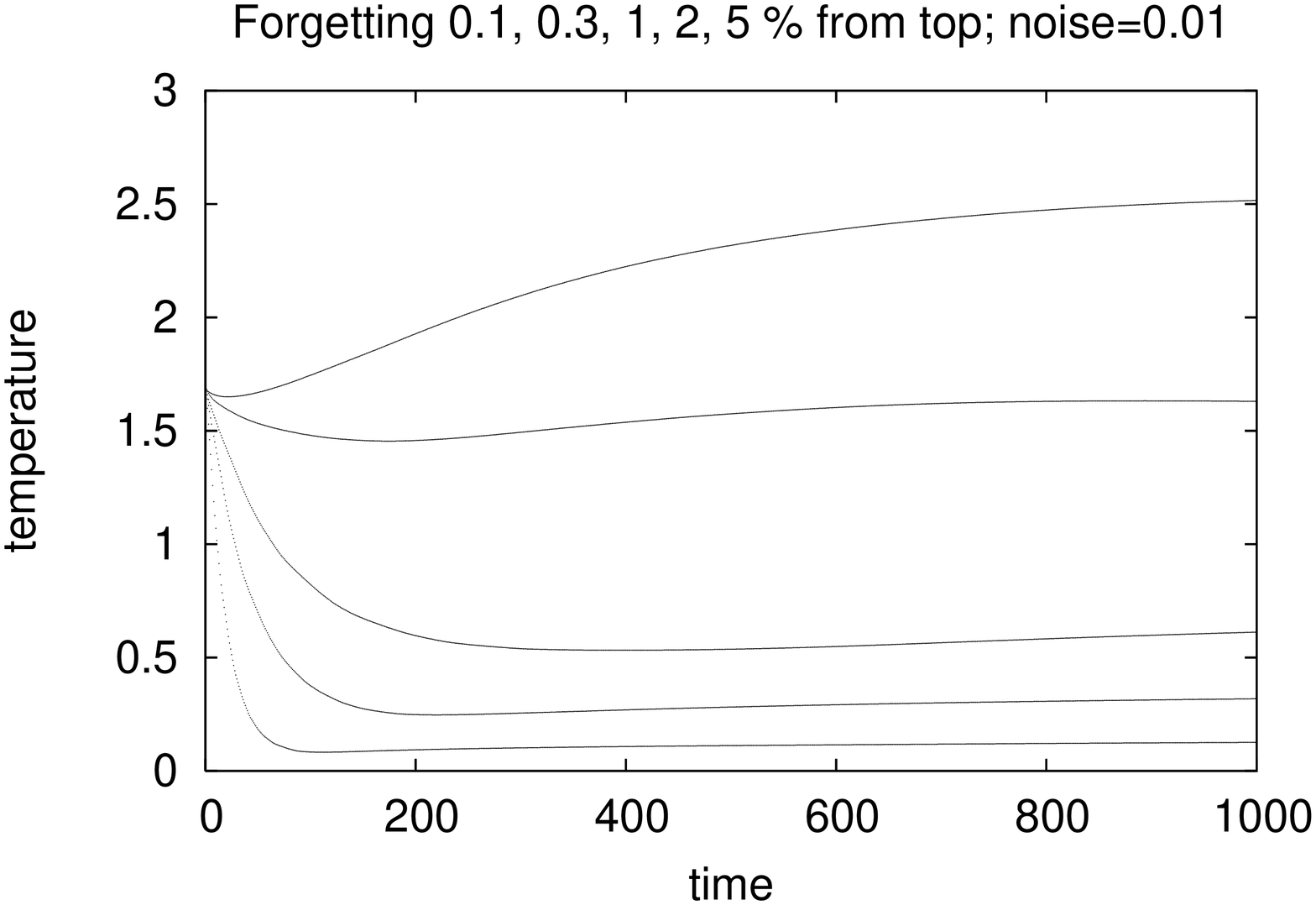}
\includegraphics [angle=-90,scale=0.3]{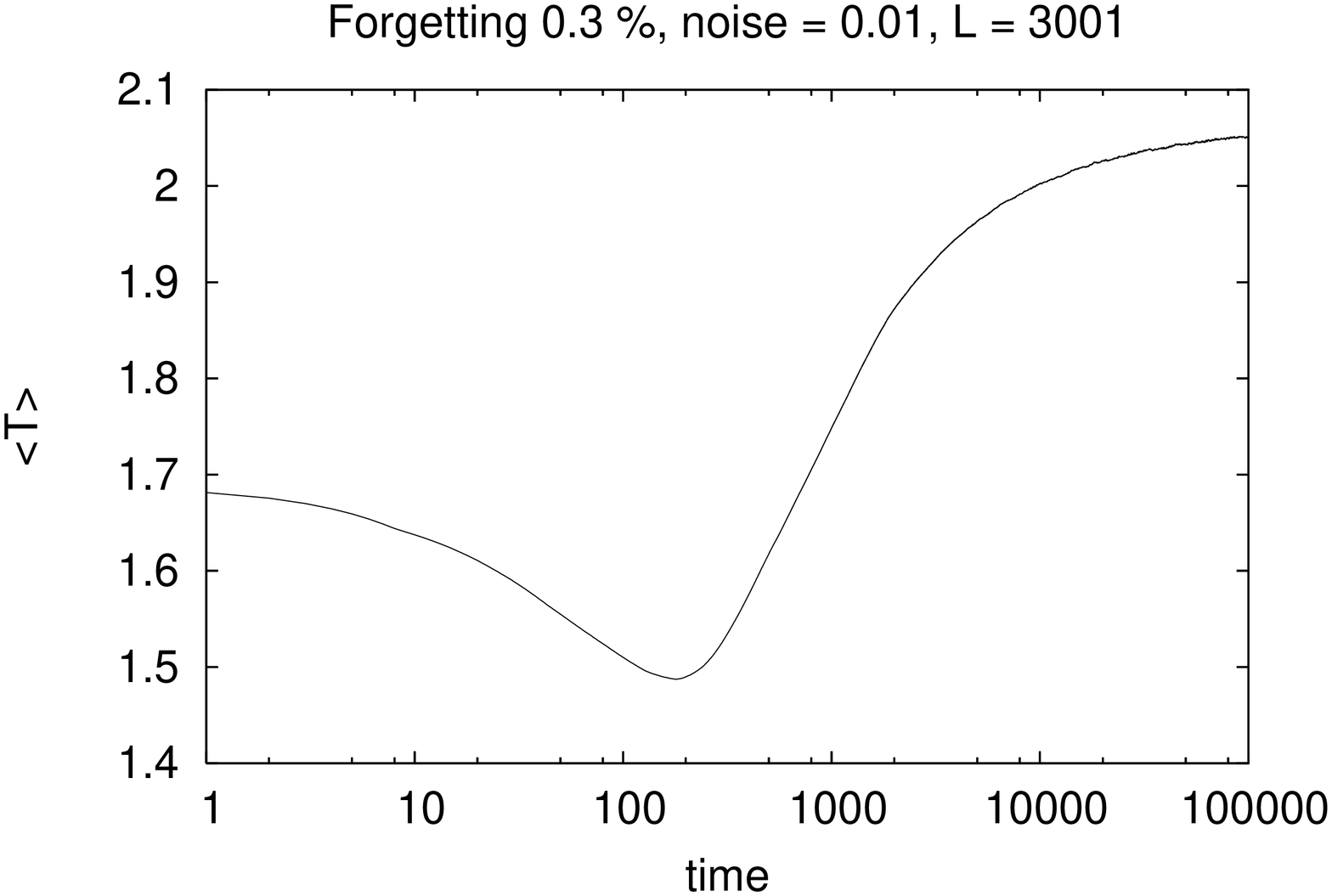}
\end{center}
\caption{Correlation and temperature $\langle T_1 \rangle$ for negligible noise 
and various forgetting rates; $L = 3001$. The lowest part shows one of the 
curves for longer times.
}
\end{figure}

\begin{figure}[hbt!]
\begin{center}
\includegraphics [angle=-90,scale=0.35]{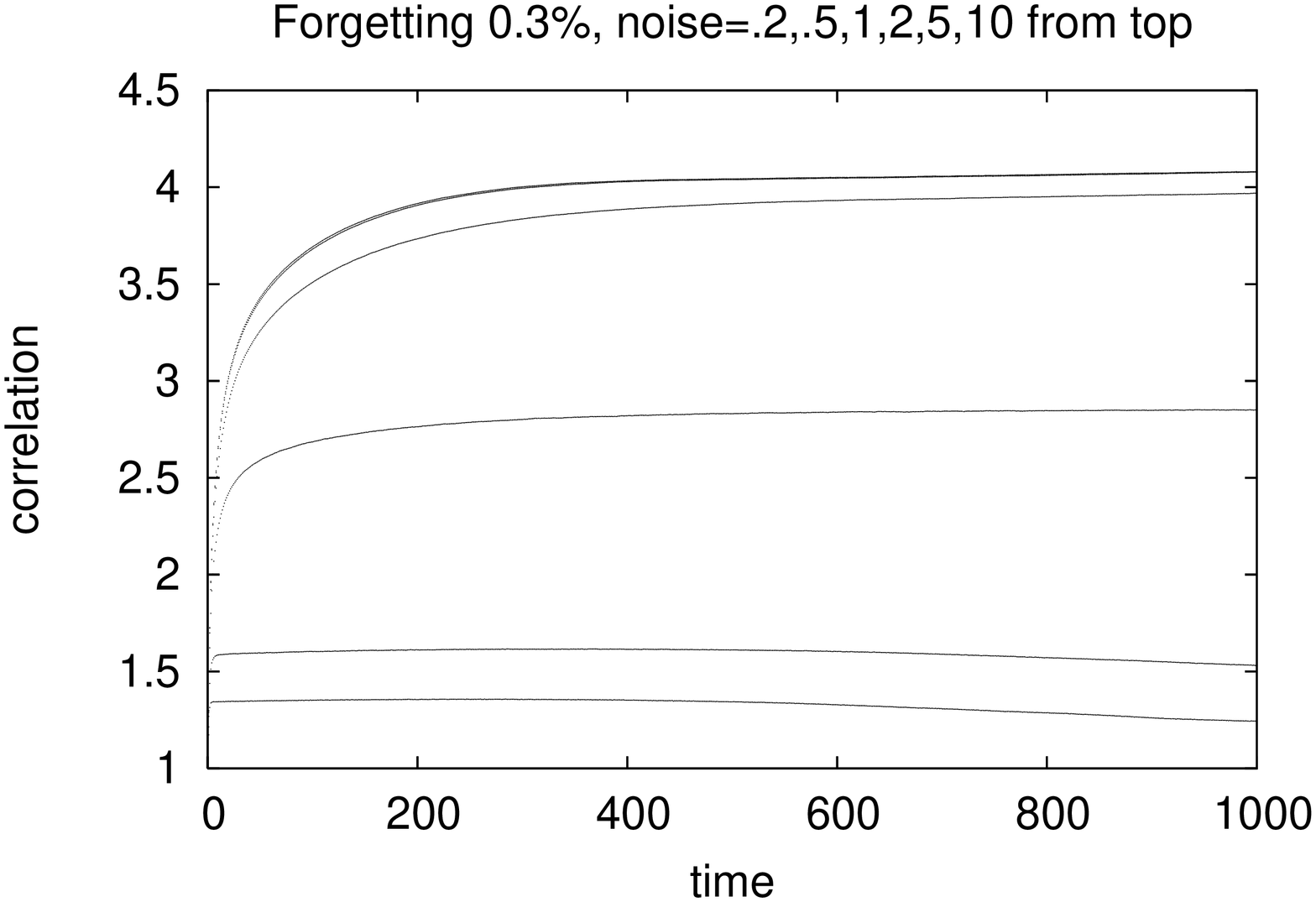}
\includegraphics [angle=-90,scale=0.35]{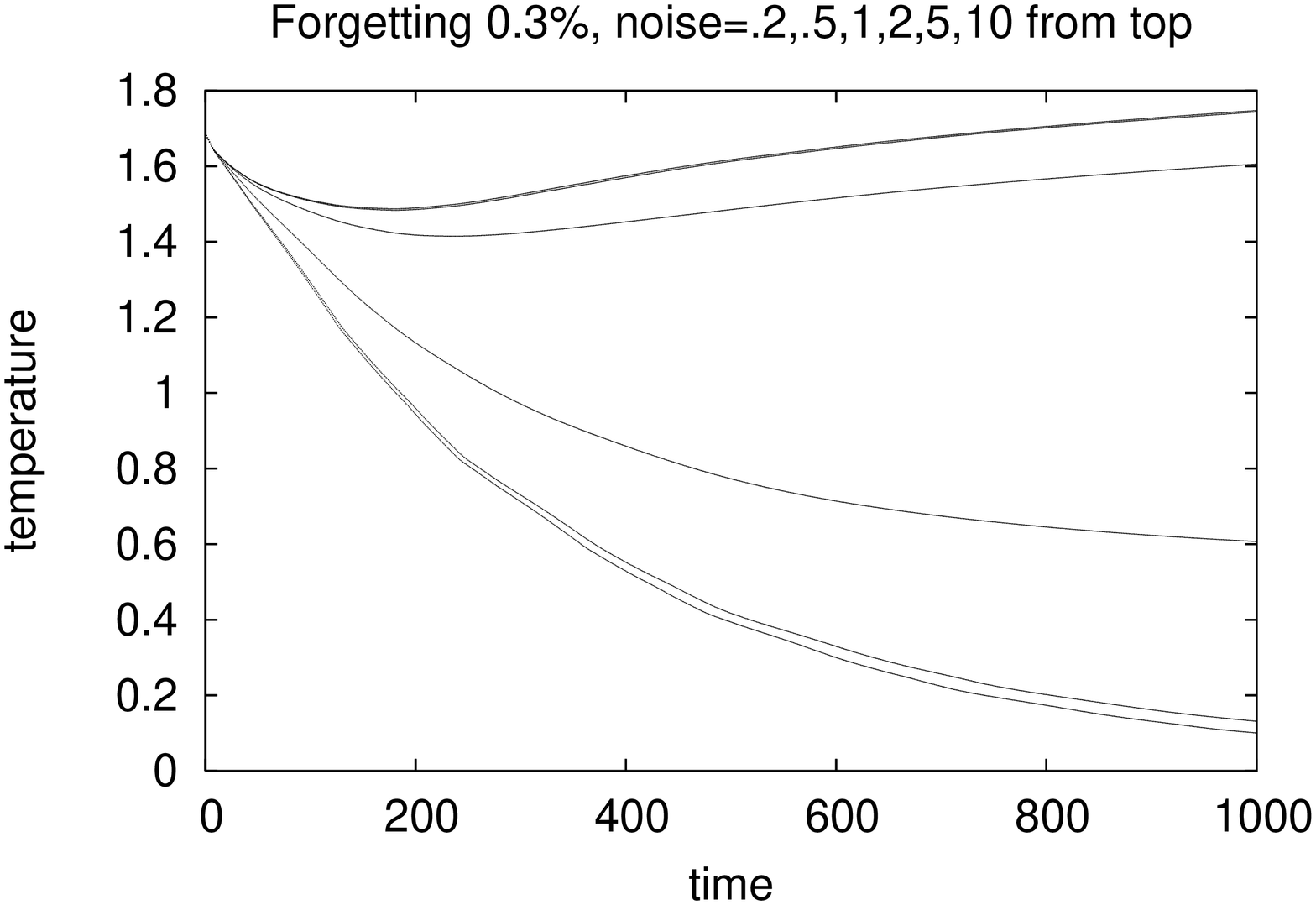}
\end{center}
\caption{As Fig. 4, for various noise levels and a forgetting rate of 0.3 \%.
The curves for $T_2=0.2$ and 0.5 nearly overlap.}
\end{figure}

\begin{figure}[hbt]
\begin{center}
\includegraphics [angle=-90,scale=0.5]{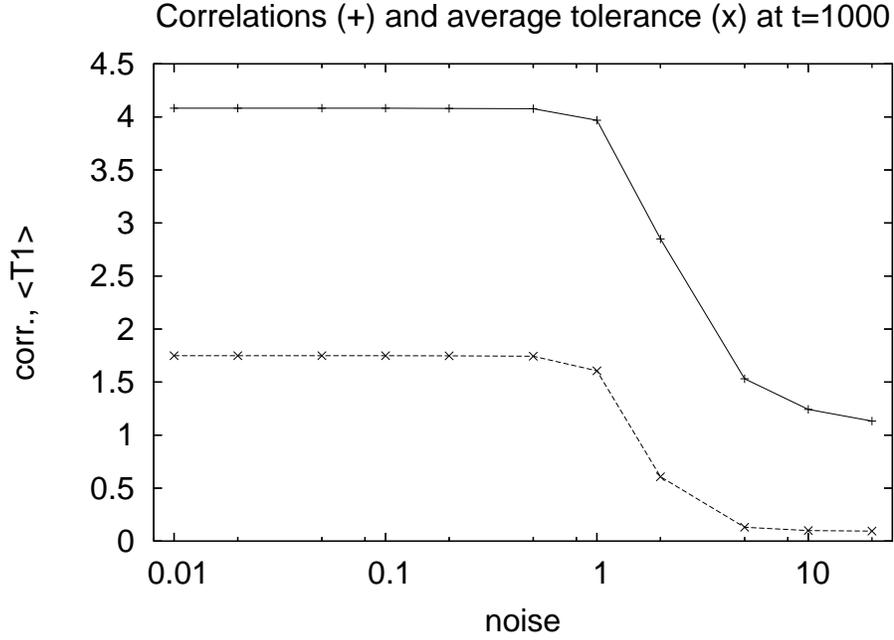}
\end{center}
\caption{Correlations at $t=1000$, from previous figure, versus noise level.
}
\end{figure}

\begin{figure}[hbt!]
\begin{center}
\includegraphics [scale=0.37]{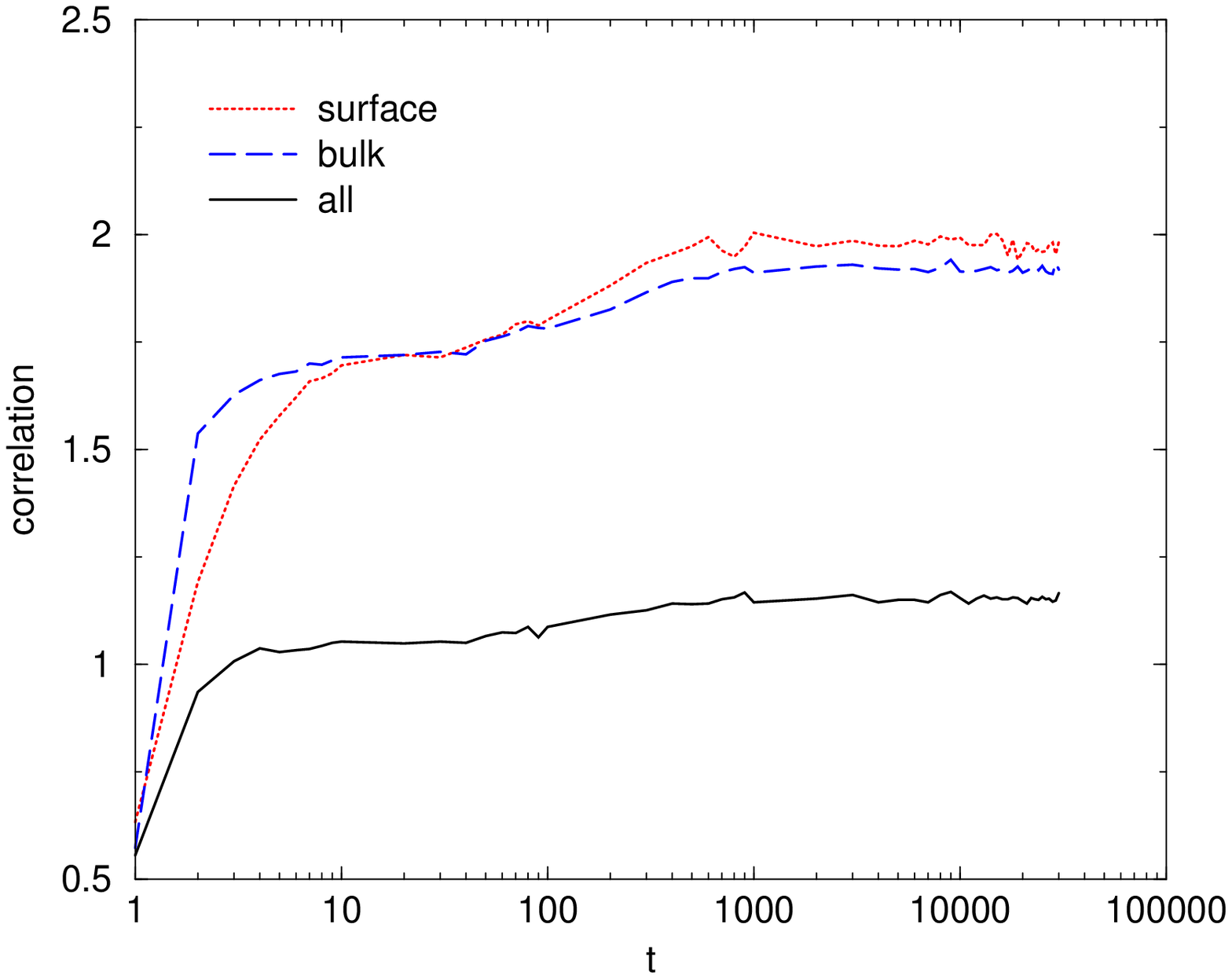}
\includegraphics [scale=0.37]{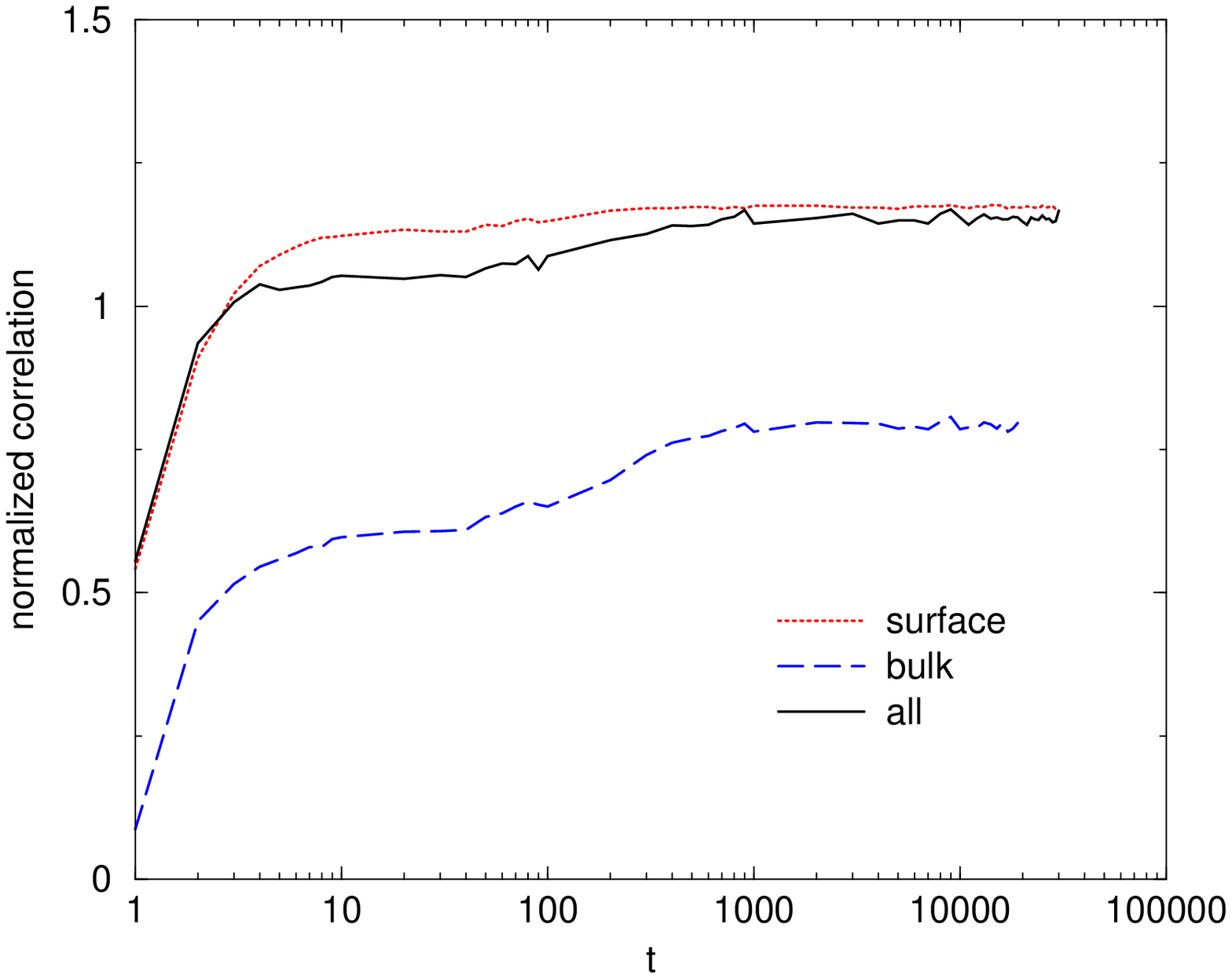}
\end{center}
\caption{Noise $k_BT_2/J=2.5$ applied to surface sites only, to bulk sites only,
and to both (left part). The right part applies a normalisation by the number of
surface or bulk sites and shows that bulk noise reduces more strongly the
segregation. 1 \% forgetting, 0.003 tolerance steps.}
\end{figure}

As promised, now we introduce a second temperature $T_2$ in addition to the 
$T_1=T$ used above \cite{odor}. At each spin flip attempt,
the spin flips if a random number is smaller than the above Glauber probability
determined by the tolerance, or if another random number is smaller than the 
Glauber probability corresponding to the noise $T_2$. Thus both flips happen 
more often when the amount of unhappiness decreases more strongly, depending 
on exp($\Delta E/k_BT_1$) or exp($\Delta E/k_BT_2$), respectively. Such 
simulations were made first by \'Odor \cite{odor} for spin 1/2 Ising models; 
noise lets the spontaneous magnetisation (= strength of long-range segregation)
jump to zero, Fig.4 below. Also for our $Q=5$ Potts-like model and strong noise 
(x symbols in Fig. 2), segregation is strongly reduced, though in a continuous 
way. Fig.3 shows better the dependence on the strength $T_2$ of the noise.

Now we allow for individual temperatures $T_1(i)$ depending on the lattice site
$i$. As discussed in the preceding section, at each iteration $T_1(i)$ 
increases by 0.01 if all four nearest neighbours belong to the same group
as $i$, and it decreases by the same amount 0.01 if none of the four neighbours
belong to the same group as $i$. In addition, at each iteration all $T_1$ 
are decreased by a small percentage, i.e. tolerance is slowly forgotten.

Fig.5 shows the resulting correlation $C$ and average temperature $T_1$ for 
negligible noise and various forgetting rates. Fig.6 shows the same two 
quantities for various noise levels at a forgetting rate of 0.3 percent. Fig.7
shows these results at the end of the simulation, $t = 1000$, against the
noise level $T_2$; for $T_2$ up to 1 the noise effects are nearly negligible,
and above $T_2 = 5$ their effects are strong and nearly independent of $T_2$.  

[For large $T_2$ we omitted the reduction of $T_1(i)$ by 0.01 since otherwise 
these temperatures became negative. Apparently for $Q > 2$ the condition
$n=0$ (none of my neighbours belongs to my group) occurs much more often
than the opposite condition $n=4$ (all my neighbours belong to my group), and
thus the previous changes by $\pm 0.01$ no longer balance each other. Thus
we used only increases by 0.01 if $n=4$, balanced by the overall forgetting
rate of 0.003.]

From \'Odor's simulations, Fig.8 distinguishes between the noise $T_2$ applied to 
the surface sites, to the bulk sites or to both. Its left part shows the lowest 
segregation if both surface and bulk sites are subjected to noise. However, its 
right part normalizes the functions by the number of surface and bulk 
sites, respectively, and then shows that bulk noise reduces segregation much
more than surface noise. (Here, ''bulk'' are the sites surrounded by four 
neighbours of the same group, and ``surface'' is the opposite case.)

Thus this section showed that a lot of external noise can perturb the personal 
preferences for neighbours of the same group. For two groups \cite{odor},
there is a first-order phase transition at some critical noise level; for
$Q = 5$ there is no such sharp transition, neither as a function of $T_1$ 
nor as a function of $T_2$, but noise still can reduce appreciably the strength 
of segregation. This sounds trivial but for zero noise the Schelling and Ising 
models do not lead to large domains \cite{jones,kirman,solomon,redner}. Thus 
only small but nonzero noise can produce segregation into ghettos. 

\section{Summary}

This review and the earlier publications from physicists 
\cite{levy,ortmanns,schulze,kirman,muller,odor} tried to overcome the 
segregation between sociology and physics with regard to the possible
self-organisation (``emergence'') of residential segregation in cities.
Besides outside forces like racial discrimination, also personal preferences
can lead to segregation, as pointed out by Schelling \cite{schelling} through
his Ising-like model. Whether this clustering leads to ``infinitely'' large 
domains (=ghettos) depends on details. The original Schelling model failed
to give ghettos; with some randomness \cite{jones}, neutral moves
\cite{kirman}, or positive temperature \cite{solomon} that can be repaired.
Nevertheless, the inclusion of empty residences and the search for the nearest
suitable residence make the Schelling model unnecessarily complicated, and
the two-dimensional Ising model seems to be a suitable simplification, giving
infinite domains for $0 < T < T_c$.

Schelling not only asked the right question but also invented a model
similar to those typically studied by physicists like the Ising model
of 1925. Thus sociologist who like the Schelling model should not complain about
physics models with up and down spins to be too unrealistic, and physicists
studying up and down spins should not claim that they bring something very
novel to sociology.

\bigskip
We thank S. Solomon, G. \'Odor, K. M\"uller, W. Jentsch,  W. Alt and S. Galam
for cooperation and suggestions.


\end{document}